\documentclass[prd, twocolumn,10pt]{revtex4-1}
\date{\today}

\newcommand{\insertplot}[5]{\begin{figure}
 \hfill\hbox to 0.05in{\vbox to #5in{\vfill
 \inputplot{#1}{#4}{#5}}\hfill}
 \hfill\vspace{-.1in}
 \caption{#2}\label{#3}
 \end{figure}}
 \newcommand{\inputplot}[3]{
 \special{ps: plotfile #1}
}
\newcounter{fig}   

\usepackage{epsfig}
\usepackage{amsmath}
\usepackage{amsfonts}
\usepackage{graphicx}
\usepackage[german, english]{babel}
\usepackage{amssymb}
\usepackage{ifthen}

\begin{document}

\newcommand{\dd}{\mbox{d}}
\newcommand{\tr}{\mbox{tr}}
\newcommand{\la}{\lambda}
\newcommand{\ta}{\theta}
\newcommand{\f}{\phi}
\newcommand{\vf}{\varphi}
\newcommand{\ka}{\kappa}
\newcommand{\al}{\alpha}
\newcommand{\ga}{\gamma}
\newcommand{\de}{\delta}
\newcommand{\si}{\sigma}
\newcommand{\bomega}{\mbox{\boldmath $\omega$}}
\newcommand{\bsi}{\mbox{\boldmath $\sigma$}}
\newcommand{\bchi}{\mbox{\boldmath $\chi$}}
\newcommand{\bal}{\mbox{\boldmath $\alpha$}}
\newcommand{\bpsi}{\mbox{\boldmath $\psi$}}
\newcommand{\brho}{\mbox{\boldmath $\varrho$}}
\newcommand{\beps}{\mbox{\boldmath $\varepsilon$}}
\newcommand{\bxi}{\mbox{\boldmath $\xi$}}
\newcommand{\bbeta}{\mbox{\boldmath $\beta$}}
\newcommand{\ee}{\end{equation}}
\newcommand{\eea}{\end{eqnarray}}
\newcommand{\be}{\begin{equation}}
\newcommand{\bea}{\begin{eqnarray}}
\newcommand{\ii}{\mbox{i}}
\newcommand{\e}{\mbox{e}}
\newcommand{\pa}{\partial}
\newcommand{\Om}{\Omega}
\newcommand{\vep}{\varepsilon}
\newcommand{\bfph}{{\bf \phi}}
\newcommand{\lm}{\lambda}
\def\theequation{\arabic{equation}}
\renewcommand{\thefootnote}{\fnsymbol{footnote}}
\newcommand{\re}[1]{(\ref{#1})}
\newcommand{\R}{{\rm I \hspace{-0.52ex} R}}
\newcommand{\N}{{\sf N\hspace*{-1.0ex}\rule{0.15ex}%
{1.3ex}\hspace*{1.0ex}}}
\newcommand{\Q}{{\sf Q\hspace*{-1.1ex}\rule{0.15ex}%
{1.5ex}\hspace*{1.1ex}}}
\newcommand{\C}{{\sf C\hspace*{-0.9ex}\rule{0.15ex}%
{1.3ex}\hspace*{0.9ex}}}
\newcommand{\eins}{1\hspace{-0.56ex}{\rm I}}
\renewcommand{\thefootnote}{\arabic{footnote}}

\title{Oscillon resonances and creation of kinks in particle collisions}

\author{T.~Romanczukiewicz$^{\dagger}$ and Ya. Shnir$^{\star}$}
\affiliation{$^{\dagger}$Institute of Physics, Jagiellonian University, Krakow, Poland\\
$^{\star}$Department of Mathematical Sciences, University of Durham, UK}

\begin{abstract}
We present a numerical study of the process of production of kink-antikink pairs in the collision of
particle-like states in the one-dimensional $\phi^4$ model.
It is shown that there are 3 steps in the process, the first step is to excite the oscillon intermediate state in
the particle collision, the second step is a resonance excitation of the oscillon by the incoming perturbations,
and finally, the soliton-antisoliton pair can be created from the resonantly excited oscillon.
It is shown that the process depends fractally on the amplitude of the perturbations and the wave number
of the perturbation. We also present the effective collective coordinate model for this process.
\end{abstract}


\maketitle


{\it Introduction.~~}
Non-linear field theories in the weak coupling regime usually contain two different
mass scales associated with the perturbative particle-like states and with the soliton sector
of the model, respectively. Conjecture about the role
nonperturbative effects, related with the production of the soliton-like states
in particle collisions, may play in high energy physics
\cite{RingEsp}, has attracted a lot of attention recently. Over the last two decades
the problem of the transition between perturbative and non-perturbative sectors of
the theory has been considered in several contexts.

Simplest example of the topological solitons in one dimension is the kink solution of the
$\phi^4$ model. This model has a number of application in condensed matter physics \cite{Solitons},
field theory \cite{MantonSutcliffe,Vachaspati:2006zz} and cosmology \cite{Vilenkin}.
Dynamical properties of kinks, the processes of their scattering, radiation and annihilation
have already been discussed in a number of papers, see e.g.
\cite{Makhankov:1978rg,Moshir:1981ja,Campbell:1983xu,Anninos:1991un,Manton:1996ex,Belova:1997bq,Kiselev:1998gg,Romanczukiewicz:2005rm}.
In integrable theories, like the sine-Gordon model, there is no energy loss to radiation and kinks do not
annihilate antikinks.
However in the non-integrable $\phi^4$ model, the radiation effects in the process of kink-antikink ($K\tilde K$)
collision
become very important and depending on the
impact velocity, the collision may produce various results, e.g., an oscillating
bound state can be formed, also the soliton and antisoliton may bounce and reflect from each other.

Although the process of annihilation of the $K\tilde K$ state of the $\phi^4$ model
has been investigated in detail \cite{Campbell:1983xu,Anninos:1991un,Belova:1997bq}, there is not
much information about the inverse process, the creation of the $K\tilde K$ pairs by the collision of
two identical bunches of particles. In the recent work \cite{Dutta:2008jt} production $K\tilde K$ pair
was considered in assumption that two colliding wave trains are composed of the bunches of identical breathers,
i.e., tightly coupled $K\tilde K$ states.
Evidently, the kink-antikink production may proceed even in the case when there is no
kink-like states in the initial configuration at all. Here we aim to elucidate the mechanism for this process.

In is known that the collision of a kink and an antikink is chaotic, i.e., for some values of the
impact velocity the solitons bounce back while for some different impact velocity,
smaller or larger, they annihilate \cite{Anninos:1991un,Vachaspati:2006zz}. This behavior is related with
a resonance effect between the oscillations of the $K\tilde K$ pair and excitation of the discrete vibrational mode
of the kink.

So we might expect the opposite process of the production of the $K\tilde K$
pairs in the collision of particles also will have similar fractal character due to resonance effect
between the oscillon created in the particle collision and the oscillation of the correlated $K\tilde K$ pair.

Here we investigate the oscillon resonance numerically. We find that this resonance excitation
plays a crucial role during the process of creation of the
$K\tilde K$ pairs in the collision of particles.
We observe furthermore the fractal structure of this process.

{\it The model.~}
We consider the standard one-dimensional $\phi^4$ theory, with two vacua $\phi = \pm 1$, defined by the
rescaled Lagrangian density
\be  \label{Lag}
L = \frac{1}{2}\partial_\mu \phi \partial^\mu \phi - \frac{1}{2}\left(\phi^2 -1\right)^2.
\ee
The perturbative sector of the model consists of small linear
perturbations around one of the vacua with the mass $m = 2$.
The static kink solution for this model interpolates between the vacua $\phi_0=-1$ and $\phi_0=1$
as $x$ increases from $-\infty$ to $\infty$:
$\phi_K(x,t) = \tanh x .
$

There is a lot of similarity between the the non-integrable
model \re{Lag} and its integrable sine-Gordon counterpart.
However, the states of the perturbative sector are different in  these theories.
Evidently, in both models there are
zero translational modes in the spectrum of the linear perturbation about the kinks but
a single $\phi^4$ kink has in addition a normalizable discrete vibrational mode
which oscillates harmonically with frequency $\omega_1 = \sqrt 3$. The continuum modes on the
kink background have higher frequencies $\omega > 2$.
Evidently, if the amplitude
of the oscillation is large enough, such a periodically expanding and contracting kink can be treated as
kink-antikink-kink bound state and this excitation can be considered as an intermediate step in the process
of creation of the $K\tilde K$ pair  on the kink background \cite{Manton:1996ex,Romanczukiewicz:2005rm}.

Another situation is related to the possibility of production of the $K\tilde K$ pairs on the trivial background.
Indeed, the linear excitation spectrum around the trivial vacuum contains the
radiation modes and within the $\phi^4$ model, the
collision of these particle-like states may produce $K\tilde K$ pairs
\cite{Carvalho:2009ac}.

Note that non-linear field theories usually contain several types of topological and non-topological excitations.
Indeed, besides the solitonic configurations there is another spatially localized non-perturbative
oscillon solution which, although unstable, are extremely long-lived \cite{Bogolyubsky:1976nx,Fodor:2008es,Gleiser:2009ys}.
The oscillon states naturally appear in various models
\cite{Copeland:1995fq,Hindmarsh:2007jb,Gleiser:2007te}.

In the  $\phi^4$ model  the oscillon solutions are almost
periodic. One can find the oscillon
numerically by solving the field equation in the Fourier series in time:
\be  \label{oscillon}
\phi = 1 + \eta_0(x) + \eta_1(x) \cos (\Omega t) + \eta_2(x) \cos (2\Omega t) + \dots
\ee
If $\Omega < m = 2$, the oscillations are below the threshold and cannot propagate as
modes of the continuum, so the oscillon remains relatively stable and the $\eta_1$ term dominates.

It was pointed out recently that an oscillation mode of the $\phi^4$ model may decay into
a $K\tilde K$ pair \cite{Dutta:2008jt}.

{\it Numerical results.~~}
The initial data used in our simulations represent two widely separated identical wave trains propagating from both
sides on the trivial background towards a collision point:
\begin{multline} \label{initial}
\!\!\!\!\!
\phi(x,t)\! =\! 1\! + C [F(x+vt)\sin(\omega t+kx) +F(x-vt)\sin(\omega t-kx)],
\end{multline}
where $k$ is the wavenumber of the incoming wave, $\omega=\sqrt{k^2+4}$
is the frequency and $v=k/\omega$ is the velocity of propagation of the
wave train. We consider the envelop of the train $F(x) =[\tanh (x-a_1)-\tanh(x-a_2)]$, also
the gaussian envelope $F(x) = e^{-\alpha(x-a_3)^2}$ was used to prove that our results
are independent of the particular choice of the initial state.
The parameters $a_1,a_2$ and $a_3$ define the length of the train and the initial separation between the trains.
Typically, we used the values $a_1=10, a_2=30, a_3=20$. The amplitude $C$ and the wavenumber $k$ are the
impact parameters, which can be changed freely.
In our numerical analysis we found that after small amplitude collisions, the two wave trains
separate and move in opposite directions and the radiation is created due to the interaction between
these trains. In the center of collision an oscillating lump remains. For small amplitudes, the frequency
of the oscillation is just a bit above the mass threshold. This indicate that the lump could be
identified with low wave-number linear excitation of the trivial vacuum.
For large amplitude collisions, the remaining lump oscillates with frequency within the mass gap, so
such a state can be identified as an oscillon.

Furthermore, for a certain range of values of the impact parameters,
$C$ and $k$, we observed creation of the $K\tilde K$ pairs.
During this process also an oscillon is create in the collision center (Fig. \ref{fig:1}). The most
important feature of this process is that in the space of parameters,
the regions of creation of the solitons and the regions where this process in not taking place, are
separated by a fractal-like boundary (Fig. \ref{fig:2}).
For finite wave  trains we do not expect this boundary would be a real fractal but some properties of
scaling are observed (Fig. \ref{fig:3}). We have measured the fractal (box) dimension to be $d= 1.770 \pm 0.011$,
which is much more than 1. In the case of the gaussian envelope we found $d= 1.865 \pm 0.007$. The interesting
peculiarity of the latter case is that the $K\tilde K$ pairs can be created even if $k=0$ (standing wave perturbation).

For certain values of impact parameters, an oscillon remaining in the collision center decays into the second
$K\tilde K$ pair. Sometimes the second pair moves even faster than the first pair, and it may
annihilate with the first one creating two moving oscillons.
We know that the process of collision  $K\tilde K$ pair also leads to fractal structure in the
velocity space \cite{Campbell:1983xu, Anninos:1991un}.
In our process, instead of creation on $K\tilde K$ pair, two oscillons could also be ejected from
the collision center and after a while they could decay into two pairs of $K\tilde K$. We observed some evidences
that these processes also yield the fractal dependency of  impact parameters. This study will be reported elsewhere.
\begin{figure}
 \begin{center}
   \includegraphics[height=8cm,angle=270,bb=93 68 535 732]{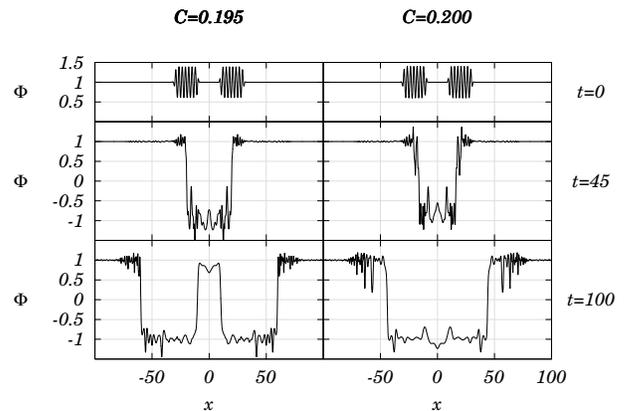}
 \caption{\label{fig:1}\small Production of the kinks in the collision of two identical wave trains.
 The initial and final field configurations are plotted at
 $t=0$, $t=45$ and $t=100$ respectively.}
\end{center}
\end{figure}

\begin{figure}
   \centering
   \includegraphics[width=8cm,bb=14 14 674 423]{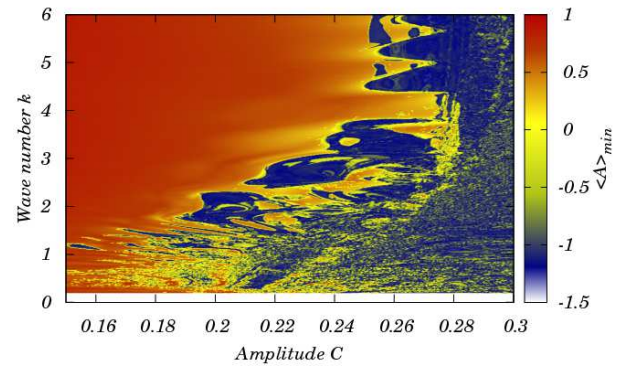}
   \caption{\label{fig:2}\small Fractal structure in the $C,k$ plane. Shading (or colour)
represent the measured minimum of average of the field
$\langle A\rangle=\frac{1}{20}\int_{-10}^{10}\!dx\; \phi(x,t)$.
The dark regions (blue in colour), where  $\langle A\rangle<-1$ indicate creation of the $K\tilde K$ pairs. }
\end{figure}
\begin{figure}
 \centering
   \includegraphics[height=8cm,angle=270,bb=66 83 551 748]{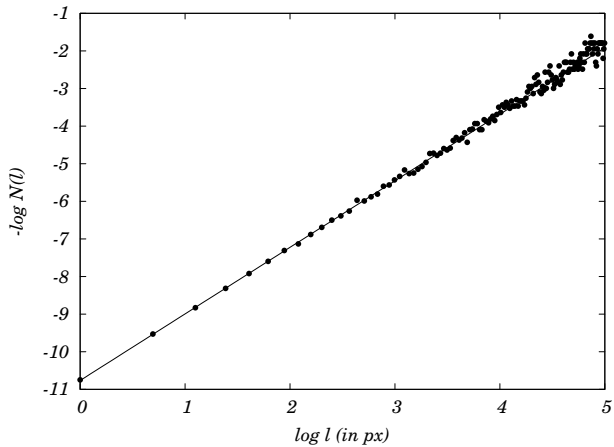}
\caption{\label{fig:3}\small Calculation of the box fractal dimension.
Number of the boxes $N$ covering the boundary is plotted versus the size of the covering boxes $l$.
The slope in the above $\log-\log$ plot gives the dimension $d= 1.770 \pm 0.011$.}
\end{figure}
To find a numerical solution of the PDE describing the evolution of the system, we used the
pseudo-spectral method. For the time stepping function we used symplectic (or geometric) integrator of
4th order to ensure that the energy is conserved.

{\it Effective collective coordinate model.~~}
In order to capture the most important steps in the process of the creation of $K\tilde K$ pair, in the collision
of two identical bunches of particles, we use the collective coordinate method which allows us to identify
the physical degrees
of freedom of the system under consideration. This approach has been applied to describe
the dynamics of the kink-antikink system \cite{Sugiyama:1979mi}.

First, we describe the process of creation of the oscillon in the collision of the incoming
wave trains.
We assume an initial field configuration on the trivial background
\be \label{incoming}
\phi(x,t) = 1 + A(t)/\cosh (x/x_0)+\xi(x,t),
\ee
which corresponds to the profile of the oscillon solution
\cite{Belova:1997bq,Fodor:2008es} with some additional perturbation $\xi(x,t)$. The gaussian
approximation to the oscillon configuration  \cite{Gleiser:2009ys} was also used to check the results.
Here the variable $A(t)$ is introduced as the collective
coordinate of the oscillon and the parameter $x_0$ represents the oscillon width. From the expansion
\re{oscillon} we know
that when $\xi=0$ the oscillon should, in the first approximation, oscillate as $A(t)=A_0\cos(\Omega t)$, where
$A_0$ is the amplitude of the oscillations, $\Omega<2$ and the value of the parameter
$x_0$ depends on the amplitude $A_0$. In the presence
of the external field $\xi$, the amplitude of the oscillon changes. However, for the sake of simplicity we set
$x_0=1.5$ as it is the width of the oscillon oscillating with amplitude $A_0=0.4$.
Substituting \re{incoming} into \re{Lag} and after integration over all space $x$
gives the effective Lagrangian which can be split into three parts: Lagrangian of the free
oscillon, Lagrangian of the perturbation $\xi$ and Lagrangian of interaction between the
oscillon and the perturbation,
\be \label{Lag-eff}
L(A,\dot A) = L_A+L_{\xi}+L_{int}.
\ee
The Lagrangian of the free oscillon has the form
\begin{equation}\label{oscillon-eff}
 L_A/x_0=(\dot A)^2-\frac23  A^4- \pi A^3-\left(4 + \frac{1}{3 x_0^2}\right) A^2.
\end{equation}
This is the Lagrangian of an anharmonic oscillator with frequency $\Omega_0 = \sqrt{4+\frac{1}{3 x_0^2}}>2$.
Since the frequency of the oscillon must be smaller than $m=2$, the amplitude of the
oscillations must be large enough to decrease the oscillation frequency below the mass threshold
\cite{Gleiser:2009ys,TRom:2009}, so the nonlinearities are crucial for the existence of the oscillon.\\
We assume that the field $\xi$, is a solution to the equation of motion of the Lagrangian $L_{\xi}$.
The perturbation $\xi$,
should represent two wave trains coming from $\pm\infty$. For the sake of simplicity we take the perturbation
of the form \re{initial}.\\
In the last part of the Lagrangian we take only the linear term in $\xi$:
\begin{equation}
 L_{int} = \alpha(t)A(t)+\beta(t)A^2+\gamma(t)\dot A+{\cal O}(A^3)+{\cal O}(\xi^2).
\end{equation}
where
\begin{equation}
 \alpha(t)=\int \!\!dx\frac{4\xi}{\cosh(x/x_0)}+\frac{\xi_x \sinh(x/x_0)}{x_0\cosh^2(x/x_0)},
\end{equation}
and
\begin{equation}
 \beta(t)=-\int \!\frac{6\xi\;dx}{\cosh^2(x/x_0)},\;\;\gamma(t)=\int\! \frac{\xi_t\;dx}{\cosh(x/x_0)},
\end{equation}
although we have also investigated the effect of the higher order terms.
Because $\xi$ is an oscillating function over a compact support, the above Lagrangian along with $L_A$ describes
dynamics which is similar to the dynamics governed by Mathieu's  equation, (especially the term proportional to
$\beta(t)$) but with additional nonlinear terms and a source term (proportional to $\alpha(t)$ and $\gamma(t)$).
We have studied the dynamics of the system numerically. Unfortunately we could not find the analytic form of
the integrals $\alpha(t)$, $\beta(t)$ and $\gamma(t)$, so we used numerical methods which are on
the same level of complexity as the explicit solution of the underlying PDE. However, within the collective coordinate
approach we can separate the most important degrees of freedom
and show that the corresponding nonlinear interaction is responsible for generation
of the fractal structure.

\begin{figure}
   \centering
   \includegraphics[height=8cm,angle=270,bb=66 83 547 759]{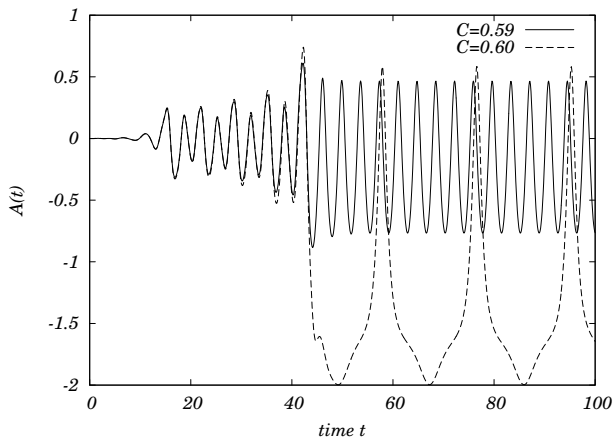}
 \caption{\label{fig:4}\small Resonance of the amplitude of the oscillon (dashed line) within the effective model.}
\end{figure}

The initial condition is that A(0)=0. As the wave train
approaches the point of the collision, the oscillon mode is excited. If the amplitude of the perturbation
is relatively small, then the oscillon, created in the
collision, oscillates with a constant amplitude around $A=0$. However, if the amplitude is large
enough, or the incoming perturbations are close to one of the (Mathieu) resonances,
the amplitude of the oscillon rapidly increases
and it starts to oscillate around $A=-1$ (or, in other words, around $\phi=0$) with amplitude of order 1, as
illustrated in Fig \ref{fig:4}. This clearly breaks
our effective approach, but it also means that the system has changed the ground state.
Such a resonant oscillation with a large amplitude, on the other hand,
shift the center of the oscillation. This transition can be related with creation of the $K\tilde K$ pairs although
the corresponding collective coordinates are not presented in our simple model \re{Lag-eff}.
Again, when we examined this effective model, we found a fractal structure on the plane $A,k$.
This fractal structure was less complicated and more localised than in the case of full PDE.
That means that although our effective model works and captures qualitatively the most important features of the full
system it also fails to reproduce some of the details, which is not a surprise for such complicated dynamical process.
We have also introduced an approximation for $\alpha(t)$, $\beta(t)$ and $\gamma(t)$ and
again, we could reproduce both the resonance excitation of the oscillon and the generation of fractal structure.

This result shows, that even after performing so many simplifications we could reproduce
(at least qualitatively) the most important features of the evolution of the system. This result confirms
our conjecture about the mechanism of the creation of the $K\tilde K$ pair in
a 3-stage process (i.e., excitation of the oscillon,
resonance and oscillon decay into the $K\tilde K$ pair).
Secondly, we conclude that the interaction between the incoming wave trains and the oscillon is the underlying
reason for the generation of the fractal structure.
Thirdly, given the generality of our approach, we expect that the effective nonlinear interactions of the same type
can be found in many different models, so the fractal structure should not be limited only
to the case of the $\phi^4$ model.

Concluding, we have found that the $K\tilde K$ pairs can be created from the
collision of particles in the 3-stage process.
In the first stage, the collision process produces an oscillon excitation.
Next, the oscillon interacts with the incoming trains, and due to the
parametric resonance it may decay into the $K\tilde K$ pair.
This process yields the fractal structure in the space of parameters of the model.
We reproduced the most
important features of the evolution of the system by constructing an effective model.
Our numerical investigations indicate fairly clearly that
the observed phenomena seems to be independent of the approximations we used.
We expect that other topological defects, like vortices, hopfions or monopoles, can be created
in the process of the same type, i.e., via resonance excitation of an oscillon created by the particle
collision.

Ya.~S. acknowledge the support from the Marie Curie Host Fellowships Programm
MTKD-CT-2006-042360 "Krakow Geometry in Mathematical Physics"

\begin{small}

\end{small}

\end{document}